\documentclass[letterpaper,twocolumn,10pt]{article}
\usepackage{usenix-2020-09}

\usepackage{tikz}
\usepackage{amsmath}

\usepackage[normalem]{ulem}
\usepackage{epsfig}
\usepackage{graphicx}
\usepackage{balance}
\usepackage{comment}
\usepackage{cite}
\usepackage{textcomp}
\usepackage{listings}
\usepackage{color,soul}
\usepackage{colortbl}
\usepackage{leading}
\usepackage{moreverb}
\usepackage{listings}
\usepackage{framed}
\usepackage{multirow}
\usepackage{wrapfig}

\usepackage{tikz}

\makeatletter
\newif\if@restonecol
\makeatother

\usepackage[boxed]{algorithm2e}
\definecolor{lightgray}{gray}{0.9}
\definecolor{lightblue}{rgb}{0.9,0.9,1}
\definecolor{red}{rgb}{1,0,0}

  {\begin{list}{$\bullet$}%
     {\setlength{\parsep}{0pt}%
      \setlength{\topsep}{0pt}%
      \setlength{\itemsep}{2pt}}}%
  {\end{list}}

\newcommand\itemno[1]{({\em #1})}

\newcommand\mathfont\mathtt

\usepackage{subfigure}
\usepackage{enumitem}
\usepackage{multirow}
\usepackage{url}
\usepackage{paralist}
\makeatletter
\g@addto@macro{\UrlBreaks}{\UrlOrds}
\makeatother

\newcommand\sname{Vronicle\xspace}
\newcommand\Sname{Vronicle\xspace}

\newcommand\os{OS\xspace}

\newcommand\contribution{Equal contribution}

\begin{document}

\date{}

\title{\Large \bf \Sname: A System for Producing Videos with Verifiable Provenance}

\author{
	{\rm Yuxin (Myles) Liu$^{\dagger}\thanks{\contribution}$, Yoshimichi Nakatsuka$^{\dagger}$$^{\textcolor{green}{\ast}}$}\\
{\rm Ardalan Amiri Sani$^{\dagger}$, Sharad Agarwal$^{\ddagger}$, Gene Tsudik$^{\dagger}$}\\
$^{\dagger}$UC Irvine, $^{\ddagger}$Microsoft\\
}

\maketitle

\begin{abstract}
Demonstrating the veracity of videos is a longstanding problem that has recently
become more urgent and acute. It is extremely hard to accurately detect manipulated videos 
using content analysis, especially in the face of subtle, yet effective, 
manipulations, such as frame rate changes or skin tone adjustments.
One prominent alternative to content analysis is to securely embed provenance information
into videos. However, prior approaches have poor performance and/or 
granularity that is too coarse. To this end, we construct \sname\ -- a video provenance
system that offers fine-grained provenance information and substantially better performance. It allows a video 
consumer to authenticate the camera that originated the video and the exact sequence of video filters that were
subsequently applied to it. \sname exploits the increasing popularity and availability of Trusted
Execution Environments (TEEs) on many types of computing platforms.
    
One contribution of \sname is the design of provenance information that allows the 
consumer to verify various aspects of the video, thereby defeating numerous fake-video creation methods.
\Sname's adversarial model allows for a powerful adversary that can manipulate
the video (e.g., in transit) and the software state outside the TEE.
Another contribution is the use of fixed-function Intel SGX enclaves to post-process videos.
This design facilitates verification of provenance information.

We present a prototype implementation of \sname (to be open sourced), which relies on current 
technologies, making it readily deployable. Our evaluation demonstrates that \sname's performance
is well-suited for off-line use-cases.
\end{abstract}

\section{Introduction}
\label{sec:intro}
Devices that can record videos surround us, ranging from standalone consumer cameras, security cameras, dashcams, webcams
and smartphones that allow users to record videos at any time. Such devices have opened up a wide variety of applications, 
most notably security-critical applications where videos are used as evidence, or the videos themselves are sensitive.
Consider the following examples:
\itemno{i} citizen journalists recording footage of an important event (e.g., a protest), or conducting an interview using smartphones,
\itemno{ii} law enforcement using body-worn cameras to record their interactions with citizens,
\itemno{iii} an electronic legal contract-signing platform using a video recording to identify the signing user~\cite{Mirzamohammadi2020},
\itemno{iv} e-voting (e.g., shareholder or political) system using videos to identify voters~\cite{Voatz}, or 
\itemno{v} video-conferencing in government, military, or corporate meetings.
Recent events have further highlighted the importance of these 
applications~\cite{Wet_signature_coronovirus, video_conferencing_coronavirus}.

Videos have been used in legal settings as evidence for a long time, as faking them was generally believed to be nearly impossible.
However, there has been an increasingly concerning trend of fraudulent videos, so-called \textit{deepfakes}, whereby an 
attacker either \textit{manipulates} an existing video or \textit{spoofs} one. One recent example is a fake video of the 
Chief-of-Staff for the prominent Russian political activist (Alexei Navalny) in a Zoom call with Dutch 
parliamentarians~\cite{deepfake_navalny_chief_of_staff}. Another example is a video of the Speaker of the US 
House of Representatives, Nancy Pelosi, which was doctored to show her struggling with 
speech~\cite{distorted_video_nancy_pelosi,distorted_video_nancy_pelosi_2}. Yet another recent example is a fake 
video used by a parent to disadvantage cheerleading rivals of her daughter~\cite{deepfake_cheerleader}.
Unfortunately, such fakes are widely shared on a variety of forums, intermingled with genuine videos, making it hard 
for the viewer to distinguish fact from fiction. The root of the problem is the the lack of any concrete means to ascertain a video's credibility.

Broadly, there are two approaches to combatting fake videos. One way is to detect them through content 
analysis~\cite{Li2018deepfake,Li2018exposing,McCloskey2018,Yang2019exposing,Korshunov2018,Rossler2019faceforensics} 
and identifying anomalies, e.g., unnatural human eye blinking. Although this approach does not impose any requirements for producing a video, 
its accuracy is low and will likely get worse as the deepfake technology becomes more sophisticated~\cite{deepfakes_havoc}. 
The aforementioned examples and others~\cite{deepfake_tom_cruise} show rapid advances in the deepfake technology.
The second approach involves embedding \textit{provenance information} into a video.
This is information about the source of the video, i.e., the actual camera device that originated it, and the sequence of all 
filters applied during post-processing. Provenance information allows a video consumer to check whether it was 
generated by an authentic camera and processed by a set of acceptable and genuine filters.

In this paper, we focus on the second approach since \textit{it has the potential to effectively mitigate fake videos} by 
providing hard-to-fabricate, detailed provenance info. Several such techniques have been recently proposed 
and they fall into two categories:
The first provides \textit{fine-grained provenance info} but achieves poor performance. For example, PhotoProof~\cite{Naveh2016} 
uses zero knowledge proofs to help authenticate transformations applied to an image. However, it takes {\bf several minutes} 
to generate a proof of transformation for a single $128\times128$ image, which is not acceptable performance for videos, 
especially, those with higher resolutions. The second category offers better performance, but provides only 
\textit{coarse-grained} provenance info. For example, AMP~\cite{england2020amp} allows video producers to sign their content, 
enabling consumers to verify the identity of the video producer. This provenance info, while useful, requires the consumer to
fully trust the video producer since there is neither a proof of the recording camera, nor of how video processing was performed 
prior to signing. Furthermore, this approach is even more ineffective for citizen journalist videos, since it only provides a 
proof of the originator, and not of the authenticity of the content.

Motivated by the need to improve upon prior techniques, we introduce \sname (\underline{V}ideo Ch\underline{ronicle}s), 
a system designed to provide \textit{fine-grained provenance info} for videos while also allowing legitimate,
\textit{fast video post-processing}.

There are three major components in the \sname architecture: \itemno{i} camera device, \itemno{ii} video post-processor, and 
\itemno{iii} video consumer. The camera device, e.g., a TEE-equipped smartphone, is responsible for capturing the video, 
generating provenance info about the device, and signing the latter along with the video.
The video post-processor receives the video from the camera device, applies the requested filters, and appends 
cryptographic proofs for each filter to the provenance info. The video consumer who receives the video and its provenance info
can check whether the video was captured on an authentic camera,
and
verify applied filters along with their parameters and the order in which they were applied.

This work answers three important research questions:

\noindent\textit{Q1. How can we construct provenance info to defeat video manipulation attacks?}
In \sname, it consists of information about: the video (e.g., timestamp, frame rate, segments, dimensions, 
and \# of frames), the camera device, and the post-processing filters. We further discuss how to verify this info.
Moreover, via a comprehensive security analysis, we show how \sname provenance info mitigates various attacks, 
including attacks on the platforms (camera or post-processor) and the video.

\noindent\textit{Q2. Can a fine-grained video provenance system achieve good performance?}
Performing video post-processing on the camera device (e.g., a typical smartphone) results in a high latency.
To this end, we exploit Intel SGX enclaves housed on powerful servers in the cloud.
Moreover, we deploy individual image filters, video decoders, and encoders in separate 
enclaves, a design called \textit{fixed-function enclaves}, in order to facilitate verification of the fine-grained provenance info.
Through extensive evaluations, we demonstrate that \sname achieves good and usable performance for offline use cases.
For example, we show that it takes about 44 seconds to upload and post-process (with 6 filters) a 10-second 720p video in \sname.
In comparison, it takes an average of 31 seconds (and maximum of 45 seconds) to upload the same video to YouTube, which (we suspect) performs some processing on it such as transcoding and copyright checks.

\noindent\textit{Q3. Can such a system be readily deployed?}
Given the immediate danger of fake videos, we believe that a viable solution needs to be immediately usable, 
without requiring any new hardware or system software solutions. \Sname instantiates a trustworthy camera 
device using Android smartphones in the form of an app secured by the recently released TEE-backed 
Google SafetyNet~\cite{safetynet}. We introduce a two-report remote attestation scheme for such a camera 
device based on the capabilities of SafetyNet in order to prevent TOCTOU attacks.
We plan to open source \sname in order to obtain feedback, foster follow-on work, and enable its adoption.
We provide a video demo of \sname.\footnote{Anonymously available at: https://rebrand.ly/vronicle}

However, we acknowedge that such camera devices remain vulnerable to certain attacks, such as kernel exploits, 
due to the limitations of SafetyNet. Therefore, we also consider another type of a camera device 
where the camera application is directly deployed within a TEE, thus providing stronger security guarantees.
While not deployable at this time, we believe that such a solution will soon be commercially available~\cite{truepic_tee_camera}.

\section{Background}
\label{sec:background}

\subsection{Trusted Execution Environments}
\label{sec:tee}
A Trusted Execution Environment (TEE) is a primitive that protects security-critical 
code and data from untrusted components through hardware-based isolation. 
A typical TEE provides the following features:

\noindent\textbf{Isolated Execution:}
A TEE creates an execution environment isolated from the rest of the system, including the \os, hypervisor, and BIOS.
Code running inside a TEE is protected against tampering during execution.
Also, data within a TEE are only available to the code running inside the same TEE.

\noindent\textbf{Attestation:}
A TEE can, upon demand, attest itself by creating a cryptographic proof of
the platform authenticity and of the integrity of the code running inside it.
Any (local or remote) party can verify this proof to determine whether it can trust that TEE and the code running within it.

One popular TEE is \emph{Intel Software Guard Extensions (SGX)}~\cite{Anati2013,Hoekstra2013,Mckeen2013}.
Isolated execution environments in SGX are called \emph{enclaves}.
A secure region in physical memory, called \emph{Enclave Page Cache (EPC)}, is reserved for enclaves.
In order to provide secure execution, functions that run inside of the 
enclave can only: (1) be called via predefined function calls, called \texttt{ECALL}s, and (2) call predefined functions outside 
the enclave, called \texttt{OCALL}s.

SGX provides two types of attestation: local and remote. Local attestation allows two enclaves that run on the same CPU 
to verify each other's identity and confirm that both are genuine SGX enclaves. Each enclave has a unique identity called 
\texttt{MRENCLAVE} -- a cryptographic hash of the code that is loaded into the enclave and other configuration details.
Using the other enclave's \texttt{MRENCLAVE} value, the attesting enclave creates an encrypted \emph{report} to prove 
that it resides on the same CPU. In remote attestation, the application enclave first performs local attestation with a special 
\emph{quoting enclave}.\footnote{Quoting enclave is assumed to run at all times.} Using an \emph{attestation key}, the latter 
generates a \emph{quote} on the report. The quote is then sent to a remote verifier.

SGX supports two types of remote attestation. The first uses Intel Attestation Service (IAS) and generates quotes using a group 
signature scheme called Enhanced Privacy ID (EPID)~\cite{Brickell2007,Brickell2009}. Since multiple group private keys are 
associated to a single group public key, a valid group signature conveys authenticity and group membership\footnote{Where 
``group'' refers to the set of all genuine SGX instances.}, while hiding the signer's identity.
This way, a verifier learns that a valid quote was generated by a genuine SGX quoting enclave without learning which
particular SGX platform hosts that enclave. The second type is called DataCenter Attestation Primitives (DCAP)~\cite{Scarlata2018}. 
It uses the EC-DSA signature scheme to generate quotes. This approach allows data centers to use their own 
quoting enclaves and a verifier can identify the exact SGX platform that generated the quote, based on the public 
key used to verify the quote. Another special enclave called \emph{Quote Verification Enclave (QvE)} is used in 
DCAP to verify quotes.

\subsection{Google SafetyNet}
\label{sec:safetynet}
Google provides a framework called SafetyNet~\cite{safetynet}, which helps developers gain trust in users' devices 
and it is used in security-critical applications such as banking. SafetyNet provides an attestation 
API~\cite{safetynet_attestation} that returns a cryptographically-signed attestation report to determine the integrity 
of the device, its system software, and the requesting app. The attested app gives a nonce to the device's Google Play Services 
to initialize the attestation process. The latter gathers necessary information from the device and sends it to Google's cloud 
service. After processing the device's information, Google's cloud service sends back a Google-signed attestation report.
Any party can verify the attestation report to find out whether the attested device has legitimate hardware, OS, 
and the claimed application.

SafetyNet helps determine whether an Android device has been rooted (e.g., by checking the presence of root binaries), 
whether the bootloader has been unlocked, whether a custom OS has been installed, and whether any rootkit has been installed.
It also attests the app package installed on the device to help determine whether the app is authentic.

In the past, it was possible to circumvent SafetyNet, e.g., using Magisk.
This was possible since SafetyNet deployed a suite of checks in the OS itself.
Recently, Google released a new version of SafetyNet, which leverages ARM TrustZone to enable hardware-backed checks.
Hardware-backed SafetyNet cannot be circumvented using the aforementioned tool~\cite{safetynet_kills_magisk}.

\subsection{Digital Video Processing}
\label{sec:vid_process}
A digital video is collection of bitmap images called \emph{frames}. 
The frame rate of a video (frames-per-second or FPS) reflects the number of frames that should be displayed in one second.

A typical raw frame is large: for example, the size of a single frame of 720p (1280 $\times$ 720) resolution is about 2.8 MB in the
RGB format and 1.4 MB in the YUV format. Therefore, to save network bandwidth, videos are encoded using efficient video 
codecs to reduce size. One of the most prominent industry standard video codecs is  
\emph{advanced video coding} (H.264), which is a lossy codec. A typical H.264-compressed video can be over a 100 times smaller than the original 
raw video, while keeping most of the detailed information.

Most videos undergo some post-processing. Modern camera devices use an Image Signal Processor (ISP) to perform 
\textit{simple} post-processing tasks, such as demosaicing and white-balancing.
Popular video editing software applications support many post-processing filters, such as repairing, enhancing, 
beautifying, special effects, image scaling, deringing, denoising, deflicking, brightness adjustment, saturation adjustment, 
contrast adjustment, sharpening, whitening, and lens flare~\cite{adobe_video_effects}.
Filters are employed for a variety of reasons, such as to undo artifacts introduced by the camera hardware to
make the final video look more ``natural'' or to blur a face for privacy.
Given the widespread use of these post-processing filters, a viable video provenance solution must support them.

\section{\sname Design Overview}
\label{sec:design}
\begin{figure}
    \centering
    \includegraphics[width=1\columnwidth]{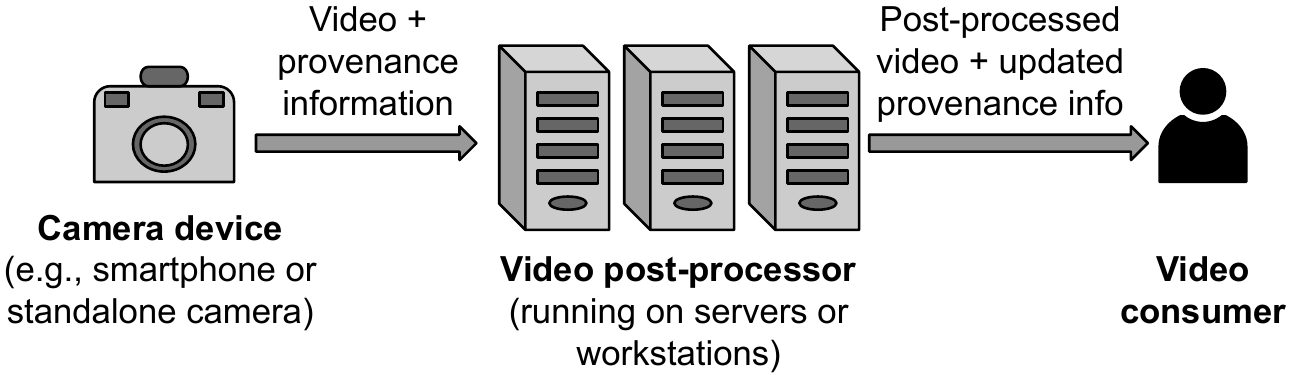}
    \caption{\em Overview of \sname.}
    \label{fig:overview}
\end{figure}

Figure~\ref{fig:overview} shows the high-level overview of \sname.
The \textit{camera device} (e.g., a smartphone or a standalone camera) generates a video.
It sends the video and some provenance info to the \textit{video post-processor}.
The post-processor applies \textit{a sequence of filters} to the video and updates the provenance info accordingly.
Finally, the \textit{video consumer} downloads the post-processed video and uses the provenance 
info to authenticate the camera that captured the video and the exact sequence of filters applied during post-processing.

\S\ref{sec:camera_realize} introduces two realizations of the camera device: one where the camera application
runs within the TEE to securely capture and sign the video, and another, where the camera application uses 
Google SafetyNet to attest the integrity of the camera device.
We define the security guarantees provided by each realization and describe and evaluate a prototype of the latter.

In our prototype, we assume the video post-processor runs entirely in the cloud for enhanced performance (see \S\ref{performance_eval} for comparison of post-processing performance in the cloud vs. on a smartphone).
However, in practice, filters can run on a smartphone, a laptop, a server, or even be split across multiple devices.
Our design assumes there is a TEE wherever the post-processor runs, and our current implementation runs on 
multiple Intel SGX enclaves in cloud VMs. Note that we use the terms ``video post-processor'' and ``server'' interchangeably.

\section{Provenance Info}
At the core of \sname is the concept of \emph{provenance info}.
Its purpose is to allow the consumer to authenticate:
(1) the actual camera device that captured the video, and
(2) the exact sequence of filters applied during post-processing.
\sname's goal is to enable the consumer to verify that the provenance info is authentic,
and has not been tampered with.

It is important to note that the consumer authenticates the applied filters, but does not verify that the applied filters were the ones requested by the operator of the camera device (i.e., the producer of the video).
This derives from our threat model (\S\ref{sec:models}), in which the consumer does not trust the operator of camera device, who might be malicious.

Next, we sketch out how provenance info is generated, verified, and used.

\subsection{Generating Provenance Info}
For the sake of simplicity and ease of presentation, for now we consider a single video frame.
Recall that a camera device is assumed to be secured directly with a TEE or indirectly through SafetyNet.
The device captures a frame, signs it, and sends it to the cloud where post-processing may be applied.
Each camera device has a unique certificate, 
which is later used by the consumer to verify the frame's origin, i.e., the camera device.
In the provenance info, the camera device encapsulates its own identity, e.g., SafetyNet remote attestation report.
To protect the integrity and authenticity of the frame ($\mathfont{F}$) and its provenance info ($\mathfont{PI}$), 
the camera device signs both using a private key, producing two digital signatures: $\mathfont{Sig_{F}}$ and $\mathfont{Sig_{PI}}$.
Then, it sends $\mathfont{F, PI, Sig_{F}, Sig_{PI}}$, and its certificate $\mathfont{Cert_{cam}}$ to the post-processor.
Note that we consider the certificate to be part of the ($\mathfont{PI}$), but highlight it separately for ease of notation.

Upon receiving data from the camera device, the post-processor applies a sequence of filters
and generates the post-processed frame, $\mathfont{F^\prime}$. It also extends $\mathfont{PI}$ to include the ordered list of 
filters and generates $\mathfont{PI^\prime}$. To prevent tampering with $\mathfont{F^\prime}$ 
and $\mathfont{PI^\prime}$, it signs these using its private key, generating $\mathfont{Sig_{F}^\prime}$ and 
$\mathfont{Sig_{PI}^\prime}$. Finally, the post-processor adds its own certificate $\mathfont{Cert_{vp}}$ to $\mathfont{PI^\prime}$.

\subsection{Verifying Provenance Info}
Upon receiving a request from a consumer, the video post-processor responds with: 
$\mathfont{F^\prime, PI^\prime, Sig_{F}^\prime, Sig_{PI}^\prime, Cert_{cam}}$, and $\mathfont{Cert_{vp}}$.
First, the consumer validates two certificates: $\mathfont{Cert_{cam}}$ and $\mathfont{Cert_{vp}}$.\footnote{
This includes multiple checks, including purely cryptographic validity, expiration, revocation,
and validation of higher-level PKI certificates.}
Then, it checks the integrity of the post-processed frame and its provenance info by verifying 
signatures $\mathfont{Sig_{F}^\prime, Sig_{PI}^\prime}$ over $\mathfont{F^\prime, PI^\prime}$.
Finally, it authenticates the camera device, checks the sequence of filters, and decides whether to accept 
the post-processed frame. Note that the consumer does not check the integrity of the original frame 
$\mathfont{F}$; we explain this in \S\ref{sec:mini_pi}.

\subsection{Supporting Videos}
\begin{figure}
    \centering
    \includegraphics[width=1\columnwidth]{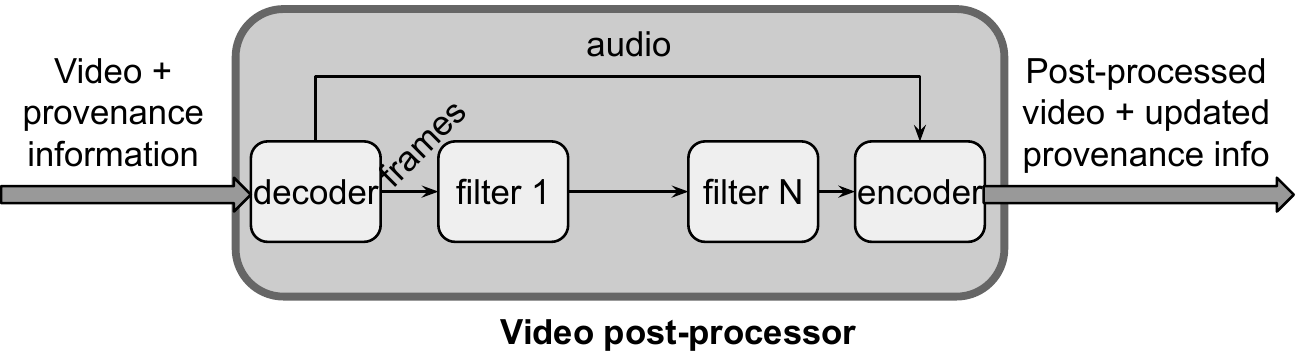}
    \caption{\em High-level view of the video post-processor.}
    \label{fig:post-processor}
\end{figure}

We now explain how \sname handles videos. As explained in \S\ref{sec:vid_process}, videos are encoded to reduce size.
This poses two challenges -- (1) image filters typically operate on raw frames; and (2) encoded formats typically include audio.
To address this, we include two additional components to the post-processor: decoder and encoder.
The former receives the video from the camera device, decodes it and forwards the video frame-by-frame to image filters.
The decoder also extracts the audio from the video.
All post-processed (filtered) frames as well as the audio are then given to the encoder, which encodes them 
into a video that is ready for the consumer. Figure~\ref{fig:post-processor} shows this design.
Note that it is possible to perform post-processing on the audio as well, although we do not currently support that in our prototype.

\subsection{Components of Provenance Info}
\begin{table*}[t] 
\centering
{\small
\begin{tabular}{|c|l|l|l|}
\hline
\textbf{General Name}                                                                            & \multicolumn{1}{c|}{\textbf{Data Name}}                  & \multicolumn{1}{c|}{\textbf{\shortstack{Data Size\\(typical /\\ bytes)}}}       & \multicolumn{1}{c|}{\textbf{Usage}}                                                                                        \\ \hline
\multirow{4}{*}{Video Info}                                                                      & Video ID                                                 & 44                                                     & Prevents swapping of frames from different videos                                                            \\ \cline{2-4} 
                                                                                                    & Timestamp                                             & 10                                                     & Prevents manipulation of date and time of video                                                              \\ \cline{2-4} 
                                                                                                    & Location (optional)                                   & 18                                                      & Helps verify where the video was captured                                                                     \\ \cline{2-4} 
                                                                                                    & Dimensions                                            & 8                                                      & Prevents manipulation of frame sizes                                                                         \\ \hline
\multirow{5}{*}{Segment Info}                                                                    & Segment ID                                               & 3                                                      & Prevents swapping of frames from different segment                                                           \\ \cline{2-4} 
                                                                                                    & Total number of segments                              & 3                                                      & Prevents segments from being dropped                                                                         \\ \cline{2-4} 
                                                                                                    & Frame rate                                            & 3                                                      & \begin{tabular}[c]{@{}l@{}}Prevents frame rate attacks\\ Necessary to encode video correctly\end{tabular}    \\ \cline{2-4} 
                                                                                                    & Total number of frame(s)                                & 3                                                      & Prevents frames from being dropped                                                                           \\ \hline
\multirow{4}{*}{Filter Info}                                                                     & Number of filter(s) (F)                                        & 3                                                      & Prevents wrong number of filter from being applied                                                            \\ \cline{2-4} 
                                                                                                   & Ordered list of applied filter(s)                      & 15 $\times$ \# of F                           & Prevents wrong order of filters being applied                                                                  \\ \cline{2-4}
												    & Identity (\texttt{MRENCLAVE}) of applied filter(s)                                                    & 44 $\times$ \# of F                                                                & Prevents wrong filters from being used                                             \\ \cline{2-4}
                                                    & Parameters (P) of applied filter(s)                                                                       & 8 $\times$ \# of P                                       & Prevents wrong filter parameters from being used                                                     \\ \hline
\multirow{3}{*}{\shortstack{Encoder/\\Decoder Info}}                                                            & Identity (\texttt{MRENCLAVE}) of encoder and decoder     & 88                                   & Prevents tampering at the decoder or encoder                                                                                    \\ \cline{2-4}
                                                    &  \begin{tabular}[c]{@{}l@{}}Frame tag (i.e., a frame ID)\\ Added by the decoder/removed by the encoder\end{tabular}    & 3                                               & Prevents reordering of frame                                                                       \\ \hline
\begin{tabular}[c]{@{}c@{}}Camera \\Device Info\end{tabular} & E.g., SafetyNet attestation reports                                                            & 10810                             & Prevents incorrect or malicious camera devices                                                                                 \\ \hline
\end{tabular}
\caption{\em Components of provenance information in \sname.}
\label{tab:metadata}
}
\end{table*}

Table~\ref{tab:metadata} shows the components of provenance info in \sname and their roles and sizes.
It includes information about the video, such as the video ID, as well as the time and location of video capture, 
provided by the camera device, although our current prototype does not support the latter yet.
This can be useful when it is important to know \textit{when} and \textit{where} content was originally captured.
It also includes information about the video segment.
We assume a video is broken into multiple segments and uploaded for post-processing segment-by-segment.
We use segments to help process a video in smaller chunks. In fact, when referring to ``a video sent from the camera device,'' 
we actually mean ``a video segment.'' Provenance info further includes the list of filters, identity of SGX enclaves that host 
each filter (\S\ref{sec:multi}), as well as the camera device info, such as a SafetyNet attestation report.

Table~\ref{tab:metadata} also shows some intermediate info, which is added by the decoder and removed by the encoder.
This is used to identify the frames passed to post-processing filters (as discussed in \S\ref{sec:fixed_tracking});
needed to defend against an attacker that can reorder or drop the frames passed to the filters (\S\ref{sec:secAnalysis}).

\subsection{Post-Processing Realization}
Video post-processing can run on any device as long as the result is deemed verifiable and trustworthy by the consumer.
We use SGX enclaves to benefit from their high-performance and highly-isolated execution environment.
Also, SGX remote attestation can be used to cryptographically bind enclave-generated public keys with 
enclave code via certificates ($\mathfont{Cert_{vp}}$), allowing consumers to verify results based on the enclave's behavior.
However, since \sname allows the use of any video filters, authenticating the enclave code is not trivial.
This is especially the case when \sname runs multiple filters, as there are numerous filter combinations.
We address this issue in \S\ref{sec:multi}.

We support both IAS-based and DCAP-based remote attestations in \sname as they provide a useful trade-off.
With DCAP, the consumer can authenticate the data center that hosted the post-processor enclave(s), which can help the consumer better evaluate the trustworthiness of the video.
With IAS, the post-processor host remains anonymous, which might be necessary in order to fully anonymise the video.

\subsection{Trusted Camera Realizations} \label{sec:camera_realize}
We now introduce two camera device realizations.
The first, hereafter referred to as \textit{SafetyNet-Camera}, uses Google SafetyNet (\S\ref{sec:safetynet}).
All of the camera device functionality is implemented as an Android app, which uses the SafetyNet API to capture an 
attestation report and include it in the provenance info. We chose to implement this camera device since 
it can be immediately used on Android smartphones.

However, SafetyNet attestation presents a particular challenge that we have to address.
SafetyNet only attests that the OS and the app are not compromised at the time of the check.
This creates an opportunity for a TOCTOU attack, whereby the attacker compromises the device (e.g., roots it) 
{\bf after the check and before} using the camera app. To mitigate this, we deploy a two-report scheme:
\textit{(i)} the camera app generates a fresh key-pair and uses the hash of the public key as a nonce 
to conduct the first round of SafetyNet attestation; \textit{(ii)} the user records a video using the camera application;
\textit{(iii)} the app again uses the hash of the public key as a nonce to conduct another round of SafetyNet attestation. 
\textit{(iv)} the app uses the private key to sign the video as well as its provenance info and then erases that key.
Besides preventing TOCTOU attacks, this scheme uses a new key-pair and new attestation reports for each video, 
thus preventing reuse.
Finally, this scheme binds the pubilc key generated for the video to the app that generated it via the SafetyNet report.

We note that SafetyNet cannot detect all types of OS compromises, such as those exploiting kernel vulnerabilities.
Therefore, we introduce a second camera realization, the \textit{TEE-camera}, in which the camera app is deployed 
within a TEE, running inside an ARM TrustZone as a trusted app. This camera device provides stronger security 
guarantees. While this approach is not ready for immediate adoption, it is likely to become commercially available soon.
For example, a startup called Truepic recently claimed that, by partnering with Qualcomm, it produced a camera application 
for capturing photos natively in Qualcomm TEE (based on ARM TrustZone) on a specific Google Pixel 
device~\cite{truepic_tee_camera}. 

\subsection{Third party verifiers}
\label{sec:third_party}
In \sname, video consumers use provenance info to decide whether to trust a video before watching it.
However, doing so is not trivial as it requires performing several checks and reasoning about filters and their parameters.
While savvy users can do this on their own, we envision a third party verifier that takes the 
provenance info, rates the trustworthiness of the video, and makes the result available to the end-users.
This is similar to how rating agencies in the financial markets rate various financial assets.
While more experienced investors can directly study each financial asset to make a decision on their own, 
most investors simply trust the ratings by well-known and reputable rating agencies.
We leave it to future work to develop an algorithm to rate the trustworthiness of a sequence of specific filters.

\section{Threat/Adversary Model} \label{sec:models}
The \sname ecosystem includes four key entities: (1) camera device, (2) video post-processor, (3) a set of SGX enclave(s) 
in the post-processor, and (4) the consumer. Since \sname's primary goal is for consumers to gain trust in videos using 
verifiable video provenance, we discuss the threat model from the consumer's perspective.

We consider two types of adversaries.
The first adversary tries to generate a fake video or tamper with a video produced by \sname.
The second adversary attacks the camera device or the post-processor.
Examples of this adversary are the owner/operator of the camera device, 
or the operator of the data center where post-processing happens.

Both types aim to produce tampered or spoofed videos 
that appear genuine to the consumer. We assume that the first adversary type can create arbitrary videos and provenance 
info, or can make arbitrary changes to a video produced by \sname, e.g., insert frames, drop frames, crop frames, or
change the frame rate. We also assume that the second adversary type can: install arbitrary apps on the camera device, 
unlock the bootloader and install a custom \os, root the device, install rootkits, and/or exploit \os kernel vulnerabilities.
In case of the SafetyNet-camera device (\S~\ref{sec:camera_realize}), which uses a TEE as well 
as in the case of the TEE-camera, we assume that the consumer can trust the TEE, the code running inside it and its 
publisher, the camera vendor's certificate, as well as the corresponding attestation report, e.g., SafetyNet attestation.
Moreover, in case of the SafetyNet-camera, we trust the camera device app code and assume that the attacker 
cannot bypass, tamper with, or spoof Google SafetyNet checks.

We assume that the second adversary type controls the post-processor platform, its \os, applications, and the network.
It can also launch arbitrary enclaves. However, Intel SGX implementation and its remote attestation functionality used in 
the post-processor is assumed to be trusted. Furthermore, the entity that generates the remote attestation report for 
the enclaves is trusted. The consumer trusts:  (1) Intel IAS servers, if IAS is used, or
(2) the data center DCAP quoting enclave implementation, if DCAP is used.

\noindent{\bf Out-of-Scope attacks} include: side-channel attacks, Return-oriented Programming (RoP) as well as other
code-reuse attacks, and physical attacks against enclaves and camera device TEEs (and the camera device app in the case of the SafetyNet-camera).
We assume that all algorithms and cryptographic primitives implemented within enclaves and other TEEs are correct and 
immune to such attacks. We also consider Denial-of-Service (DoS) attacks to be irrelevant to this paper,
since they have nothing to do with the adverary's goal of producing fake videos.
Finally, since \sname is primarily concerned with veracity of videos (i.e., their authenticity and integrity), 
 confidentiality and secure distribution of videos (i.e., digital rights management) are out of scope.

\section{Post-Processing w/ Fixed-Function Enclaves}
\label{sec:multi}
A simple design for the video post-processor in the SGX context is to deploy it within one enclave,
which hosts all three components:  decoder, filter(s), and encoder. However, this presents two important challenges:

{\bf First}, it is hard for the consumer to verify that the correct filter binaries are deployed in the enclave 
and that the filters were applied in the correct order. This is because of many possible filters and numerous 
ways of ordering them. Since enclave remote attestation measures the code loaded at initialization time, 
we cannot allow dynamic code loading (otherwise the attacker can run arbitrary code in the enclave without being detected).
Therefore, a different enclave binary needs to be generated given a
desired set of filters and their ordering. This would complicate verification, since the consumer would need 
to re-create the binary and compare its measurement with that in the SGX remote attestation report.
{\bf Second}, it requires generating the enclave binary upon receiving the post-processing request, thus increasing 
startup latency.

\begin{figure}
    \centering
    \includegraphics[width=1\columnwidth]{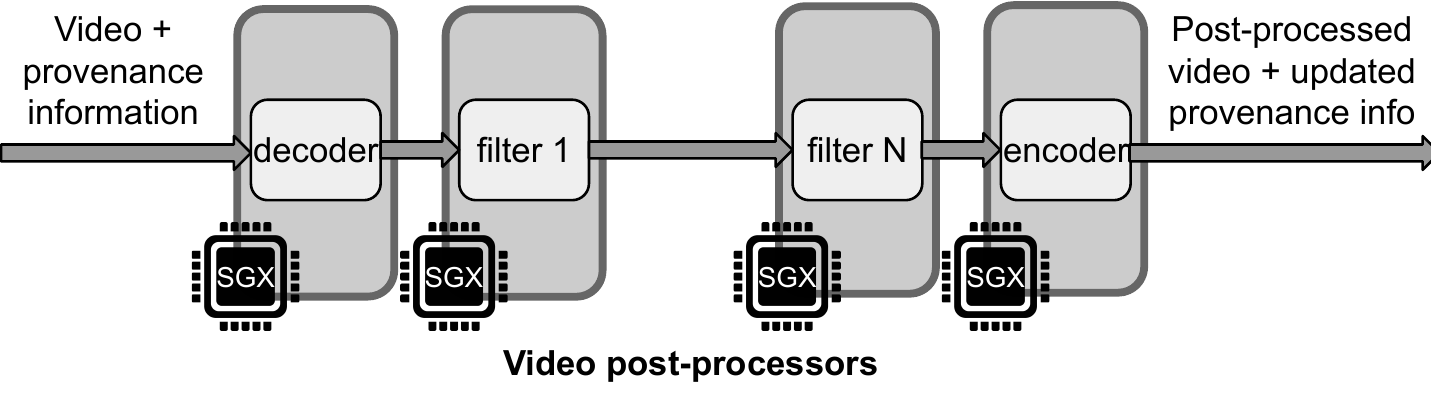}
    \caption{\em Fixed-function enclaves for video post-processing.}
    \label{fig:fixed_function}
\end{figure}

To address these challenges, we design the video post-processor using \textit{fixed-function enclaves}.
Separate enclaves are used for each filter, for the video decoder, and for the encoder.
These fixed-function enclaves can post-process the video, each appending its own remote attestation 
report to the provenance info. Figure~\ref{fig:fixed_function} shows how \sname uses fixed-function enclaves 
for post-processing. First, the video sent from the camera device is delivered to the decoder enclave.
Decoded frames are then forwarded to the sequence of appropriate filter enclaves.
The output of the last filter is sent to the encoder enclave, which generates the final post-processed video.

This design addresses the aforementioned challenges.
(1) it simplifies verification of remote attestation reports. The consumer only needs to keep a table of measurements 
for all common filters and decoders/encoders. It can then use this table to check enclave measurements in the report.
(2) it relieves the post-processor from having to generate custom enclave binaries for each video, which reduces 
the overall post-processing latency.

In addition, the fixed-function enclave design provides an additional benefit.
It enables scaling out the execution of the post-processor to multiple servers, which can help with performance under heavy load.
Note that this is not feasible if all the post-processor functions are in the same enclave.
We do not, however, explore this aspect of \sname in this paper and leave it to future work.

Nonetheless, despite its benefits, our use of fixed-function enclaves raises some issues that we address next.

\subsection{Minimizing Provenance Info}
\label{sec:mini_pi}
The discussion of provenance info in \S\ref{sec:design} does not address how the consumer can verify the
integrity of a frame provided to the post-processor ($\mathfont{F}$). Doing so requires providing the input frame 
and its signature to the consumer, which can double the bandwidth needed to retrieve a video.
The use of fixed-function enclaves exacerbates this problem, since the consumer needs to verify integrity of 
intermediate data (frames and provenance info) passed to each enclave.
We address this problem using \textit{chain verification}, i.e., we rely on each enclave to verify integrity of its input.
For example, assuming two enclaves, $\mathfont{A}$ and $\mathfont{B}$:
\begin{compactenum}
\item $\mathfont{A}$ receives data from the camera device and verifies the latter's signature.
\item $\mathfont{B}$ verifies $\mathfont{A}$'s signature on the data it receives as input, i.e., $\mathfont{A}$'s output.
\item The consumer verifies $\mathfont{B}$'s signature on the final video, i.e., $\mathfont{B}$'s output.
\end{compactenum}
This frees the consumer from checking integrity of data passed from the camera device to enclaves and between enclaves, 
hence minimizing required bandwidth. Also, since the consumer authenticates the code running in each enclave via 
its remote attestation report, it can ensure that each enclave does verify integrity of its own input.

\subsection{Flow of Provenance Across Enclaves}
\label{sec:fixed_tracking}
\begin{figure}
    \centering
    \includegraphics[width=1\columnwidth]{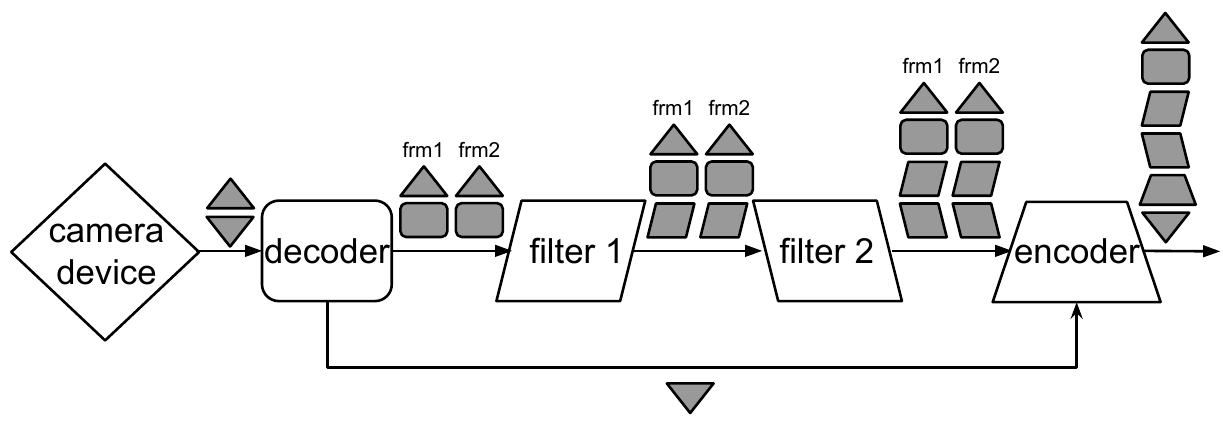}
	\caption{\em Flow of provenance info between enclaves. We assume a video consisting of two frames. Shaded boxes 
	highlight different parts of the provenance info. The shape of a shaded box generated by a module is the same as the 
	shape of that module (except for that generated by the camera device, which is split into two by the decoder). 
	``frm1'' and ``frm2'' represent frame tags added by the decoder to the provenance of each frame.}
    \label{fig:provenance_flow}
\end{figure}
While the consumer can establish trust in the enclaves via remote attestation reports, it cannot trust anything outside the enclaves.
Moreover, the fixed-function enclave paradigm requires the video to be split to frames, which go through filter enclaves independently, 
before getting merged back into the final video in the encoder. This complicates consumer verification of: (1) all frames in the 
input video being post-processed and included in the final video, (2) preservation of frame ordering in the final video, and 
(3) completeness and accuracy of the provenance info of the post-processed video. To address this, we carefully design 
the flow of provenance info between enclaves, which involves splitting, appending, and merging of provenance info 
in each enclave. Figure~\ref{fig:provenance_flow} illustrates this flow for a video with two frames processed by two filters.

\noindent\textbf{Provenance splitting.}~~
The decoder receives the video and its provenance info. After decoding the video segment, it sends a part of the video 
provenance info (along with some new info) with each of the frame to filter enclaves.
The part of the provenance info that is passed to the filter enclaves is needed to securely filter 
and encode the video segment in the encoder. It includes the video ID and segment ID, which help the 
encoder ensure that it is encoding the frames of the same video segment, as well as the total number of frames, 
which is used by filter enclaves for sanity checking. Other components of the provenance info can then be sent 
directly to the encoder. This is especially important for large components of the provenance info, mainly the SafetyNet 
attestation reports, since otherwise sending them alongside each frame would adversely affect performance.
(Smaller components of the provenance info can be optionally added to each frame provenance without impacting performance.)
Also, the decoder adds a frame-specific tag (\textit{frame id}) to each frame's provenance. This 
helps the encoder to encode the frames in the correct order.

\noindent\textbf{Provenance appending.}~~
After processing a frame, a filter enclave appends 
its \textit{MRENCLAVE} value to the frame provenance info.
This helps the consumer of the video authenticate the filter.
Moreover, the decoder and encoder enclaves 
also add their own \textit{MRENCLAVE} values to provenance info. This is needed by the consumer to verify 
that the video was correctly decoded and subsequently encoded.

\noindent\textbf{Provenance merging.}~~
The encoder receives post-processed frames along with their provenance info.
It verifies the provenance info of each frame by ensuring that their common parts (all except {\it frame ID}) match.
The verification also involves checking the  {\it frame ID} and comparing it with the total number of frames for correct and complete ordering.
The encoder then merges frame provenance into finalized provenance info for the encoded video, which includes the provenance info directly sent from the decoder. 

\section{Implementation}
\label{sec:implementation}
This section provides a detailed description of the \sname prototype.

\subsection{Camera Device}
We build an Android application for recording and uploading videos. As discussed in \S~\ref{sec:camera_realize}, we use 
Google SafetyNet to secure the application and perform two rounds of remote attestation: before and after capturing a video.
The application also allows the user to choose the set of filters to be applied.

\subsection{Consumer}
\label{ssec:consumer_application}
We build a consumer-facing video application. After downloading the video, the application verifies the 
signature and checks that provenance info contains the identity (i.e., \textit{MRENCLAVE} value) of the pre-approved 
enclaves: decoder, filters, and encoder. It also checks the SafetyNet attestation reports to authenticate the camera device.
The verification status is displayed in real time to the user. After successful verification, the application plays the video.
The video playback component is based on the open source VLC video player~\cite{vlc_website}.

Note that the consumer can not be expected to get involved in evaluating trustworthiness of filters.
Therefore, for the time being, the prototype simply checks whether each filter is in the list of pre-approved filters.
As discussed in \S\ref{sec:third_party}, rating video trustworthiness (which includes reasoning about applied filters) 
is part of our future work.

\subsection{Video Post-Processor}
The post-processor prototype has four components: H.264-based decoder and encoder, filters, and scheduler.
Each component is equipped with a TCP-based communication module, and a remote attestation module
which supports both IAS and DCAP.

The decoder, filters, and encoder each have a trusted part and an untrusted part.
The former implements a set of \textit{trusted API calls} that run within the enclave and can be called by the untrusted part.
Due to space limitations, we do not detail the roles of the untrusted part of each enclave.

Next, we discuss the roles of each enclave. We omit the flow of provenance info, since it was
already discussed in \S~\ref{sec:fixed_tracking}.

\subsubsection{Decoder}
We integrate \textit{FFmpeg}'s~\cite{ffmpeg_h264_dec} H.264 decoder to maximize our compatibilities with various 
H.264 profiles. We also integrate an open-source demultiplexer to extract H.264 stream and audio stream 
from an MP4 container.

The decoder enclave is first invoked by the untrusted part of the decoder during the remote attestation process to 
create the enclave's unique key-pair and a certificate 
and to prepare and initialize the decoder. 
In doing so, it validates the camera device certificate and uses the corresponding public key to verify the data signature. 
It also demultiplexes the MP4 container and stores a copy of the H.264 stream and its provenance info inside the enclave.
Afterwards, it only uses this secure copy to avoid TOCTOU attacks. It also signs the buffer containing 
both the audio stream and its metadata before passing them as output to the untrusted part.
Finally, the enclave is invoked to decode the frames (one call per frame) using the enclave's copy.
It also signs the buffer containing both a decoded frame and its provenance before 
passing them as output to the untrusted part.

\subsubsection{Filters}
We build a filter framework that allows any popular open-source filters to be easily integrated into \sname.
To demonstrate this, we integrate 6 popular filters: \emph{Blur}, \emph{Sharpen}, \emph{Brightness}, \emph{Grayscale}, \emph{Auto de-noise}, 
and \emph{Auto white balance}. 

Similar to the decoder, a filter enclave is first invoked by the untrusted part of the filter during remote attestation to create the enclave keys and certificate.
It is then invoked to verify the previous (decoder or another filter) enclave certificate. 
After verification, it stores the certificate inside the enclave, where this certificate is used to verify every incoming frame's signature.
Finally, it is called to apply the filter (to one frame per call).
When called, it signs the processed frame and its provenance and returns them to the untrusted part.

\subsubsection{Encoder}
We integrate both an H.264 encoder and an open-source multiplexer, where processed frames are encoded 
and multiplexed with the audio stream to an MP4 container.

Similarly to the filter enclave, the encoder enclave is first called by the untrusted part of the encoder during remote attestation to create the enclave keys and certificate and to verify the certificate of the last enclave.
It is then invoked to prepare and initialize the encoder. It requires the first post-processed frame of the video as input.
Next, it is called to encode the frames (one frame per call, where a frame, its provenance info, and their signature are passed as arguments to the call).
After verifying the signature and provenance, the enclave encodes the frame.
Also, both encoded H.264 stream and its provenance are stored inside the enclave all the time.
Finally, the enclave is invoked to multiplex both H.264 stream and audio stream into the final MP4 video container.
It takes audio stream, its metadata, their signature, and the decoder enclave certificate as arguments.
After verifying the certificate and the signature and making sure all processed frames are encoded, 
the enclave multiplexes the H.264 stream and audio stream into the final MP4 container.
It signs both the final video as well as the final provenance info and copies them to outside the enclave.

\subsubsection{Scheduler}
The scheduler is completely untrusted. Its main role is to manage the execution of other components.
Based on experiments with the prototype, we observed glitches in the latency of IAS remote attestation.
To remedy this (i.e., to obtain more predictable performance), the scheduler maintains a pool of idle decoder 
enclaves in the server. However, it does not keep a pool of idle filters and encoders, since almost all of the 
the remote attestation latency for these enclaves can be ``hidden'' within the video decoding time.

\section{Security Analysis} \label{sec:secAnalysis}

\subsection{Video Attacks}
\label{sec:integrityAttack}
\noindent\textbf{Frame Deletion:}~~
An attacker may attempt to remove certain frames from a video segment.
However, since the provenance info includes the \emph{total number of frames in a video segment},
the encoder will not produce the final video until and unless it receives the exact number of frames.

\noindent\textbf{Frame Substitution:}~~
An attacker can try to replace some frames in a video segment, e.g., using frames from a different video produced
by a different camera device, though with the same frame ID and total number of frames.
Because \emph{video ID} in the provenance info would not match that in substituted frames\footnote{Video ID is the 
hash of the camera device public key, including the modulo and public exponent.}, this attack is easily detected.
Even if an attacker modifies provenance info to match the video ID, the signature of provenance info would be
unverifiable. 

\noindent\textbf{Frame Cropping:}~~
An attacker might try to prevent the consumer from viewing a certain part of the frame by cropping it 
and falsely declaring the frame as having smaller dimensions. Since provenance info includes frame 
dimensions, this manipulation is detectable. The signature of provenance info also prevents an attacker 
from modifying frame dimensions. Alternatively, an attacker could crop a certain set of frames within a video segment.
Since filter enclaves do not accept any malformed frames or provenance info, such frames would be dropped, and the encoder will not produce the resulting video, as in the frame deletion case.

\noindent\textbf{Video Segment Omission:}~~
An attacker may try to delete an entire video segment. Once again, since the provenance info includes the 
total number of video segments and segment ID, this attack is trivially detectable.

\noindent\textbf{Video Segment Substitution:}~~
An attacker may try to substitute an entire segment. As the provenance info includes a segment ID, 
this attack would not work.

\noindent\textbf{Frame Rate Manipulation:}~~
This attack type has been observed {\em in the wild}, as mentioned in \S\ref{sec:intro}: on May of 2019, several 
news outlets reported that a distorted video of the US Speaker of the House Nancy Pelosi was being shared 
on popular social media platforms~\cite{distorted_video_nancy_pelosi}. This video showed the Speaker 
struggling with words, while in reality the frame rate of the video segment (where she was talking) was 
decreased by 75\%. \Sname detects this attack type, because the frame rate is part of provenance info.
Altering it invalidates the signature.

\subsection{Attacks on Video Provenance}
\label{vpAttack}
\noindent\textbf{Using Undesired Filters:}~~
A malicious server may try to use filters that are not included in provenance info. 
Recall that every filter enclave appends its \texttt{MRENCLAVE} value to the provenance info in the 
\texttt{Filter Info} field. Since \texttt{MRENLAVE} values are unique, they can be used to 
verify which filter was applied to the frame. Also, any modification of the \texttt{Filter Info} field in the 
metadata is easily detected via signature verification.\footnote{This signature is produced by a private key
which can be verified using a certificate signed by Intel's own private key.}

\noindent\textbf{Wrong-Order Filtering:}~~
Applying the same set of filters in a different order usually yields a different outcome.
A malicious server may try to apply filters in an order different from that declared in provenance info.
Since provenance info (covered by the signature) includes the exact order in which filters were applied,
wrong-order filtering is easily detected.

\noindent\textbf{Changing Video Origin:}~~
A malicious camera device may claim to be from a wrong vendor.
Spoofing certificates of camera vendors is infeasible because we trust the underlying cryptographic primitives to be secure.
Alternatively, a malicious server could change the video's origin in order to trick the consumer into believing that it came from a trustworthy source.
This attack is trivially detected since it would invalidate the signature of provenance info.

\subsection{Platform Attacks}
We now consider attacks on the camera device and the post-processor (refer to \S\ref{sec:models} for \sname{'s} trust assumptions).

To compromise the camera device, the attacker could: (1) modify the camera app, (2) root the device, 
(3) install a custom \os image, (4) install a rootkit, (5) bypass the attestation framework, e.g., SafetyNet attestation, 
(6) exploit a kernel vulnerability, (7) mount a code reuse attack on the kernel, (8) mount a physical attack to extract keys, 
(9) use a side-channel attack to extract keys, (10) exploit a vulnerability in the TEE, and (11) mount a code reuse attack on the TEE.
Our SafetyNet-camera protects against attacks (1)--(5). Our TEE-camera further protects against attacks (6) and (7).
Although both camera types are susceptible to attacks (8)--(11), they are out of scope of \sname threat model.

To compromise the post-processor, the attacker might: (1) modify enclave code, (2) bypass enclave attestation framework, 
(3) exploit enclave code vulnerabilities, (4) mount a code reuse attack on enclave code~\cite{Lee2017,Biondo2018}, or 
(5) use a side-channel attack to extract enclave keys.
Our post-processor prototype protects against (1) and (2), while (3)--(5) are out of the scope of \sname threat model.

Moreover, out-of-scope attacks can be addressed orthogonally.
For example, if future SGX versions eliminate currently present 
side-channels~\cite{Costan2016_2,Shih2017,hunt2018}, \sname would be 
automatically hardened against attack (5) on the post-processor.
Also, vulnerabilities in enclave code and camera TEE can be addressed by
using memory-safely vulnerability checking tools (e.g., TEEREX~\cite{Cloosters2020}) 
or by using an SDK written in a secure programming language (e.g., Rust~\cite{Wang2019}).

\section{Performance Evaluation}
This section discusses perfomance results obtained from experiments with the \sname prototype.

\subsection{Setup}
\noindent\textbf{Camera device and consumer.}~~
We instantiate the consumer on a machine with a 3.4GHz Intel Core i7-3770 4-core processor and 16GB of RAM with wired internet connection.
This machine runs Ubuntu 20.04 LTS OS with the generic Linux kernel 5.4.0. The camera device is a Samsung 
Galaxy S20+ smartphone, equipped with a Qualcomm Snapdragon 865 processor and 12GB RAM with 5GHz Wi-Fi internet connection. 
It runs Android 11 OS, and supports the TEE-backed SafetyNet Attestation API.

\noindent\textbf{Post-processor.}~~
We realize the \sname post-processor on a Microsoft Azure Confidential Computing VM. It has a 3.7GHz Intel Xeon E-2288G 
8-core processor with 32GB of RAM, and runs Ubuntu 18.04 LTS OS with Azure Linux kernel 5.4.0.

\noindent\textbf{Configurations.}~~
We use the camera device stock camera application to 
pre-record a 10-second 1080p video with 334 frames. We resize the footage into the following resolutions: 
176 $\times$ 144 (144p), 320 $\times$ 240 (240p), 640 $\times$ 480 (480p), 1280 $\times$ 720 (720p), 
and 1920 $\times$ 1080 (1080p). As default, we use a 2-second video segment (60 frames) with a 720p resolution, 
the blur filter, and IAS-based remote attestation. 
We ran each experiment at least 3 times. 

The default setting for each filter is as follows:  Convolution matrix of 7 $\times$ 7 for both the blur and sharpen filters 
and 0.2 decrease for the brightness filter. All other filters (gray, de-noise, and white balance) have no configuration parameters.
We partition measured latency into three parts: uploading, processing, and downloading.
Uploading represents the time from when the scheduler establishes a connection with the camera client to 
when the scheduler finishes receiving all files; a small portion of enclave preparation is also counted in this,
since we utilize multiple threads. Processing represents the time from when the scheduler finishes receiving all  
files to when the encoder finishes encoding the last frame. Downloading represents the time between the encoder establishing 
a connection with the viewer client and the encoder getting the confirmation from the client that it has successfully 
received all files. 

\subsection{Evaluation Results}
\label{performance_eval}
\begin{figure}
    \centering
    \includegraphics[width=0.8\columnwidth]{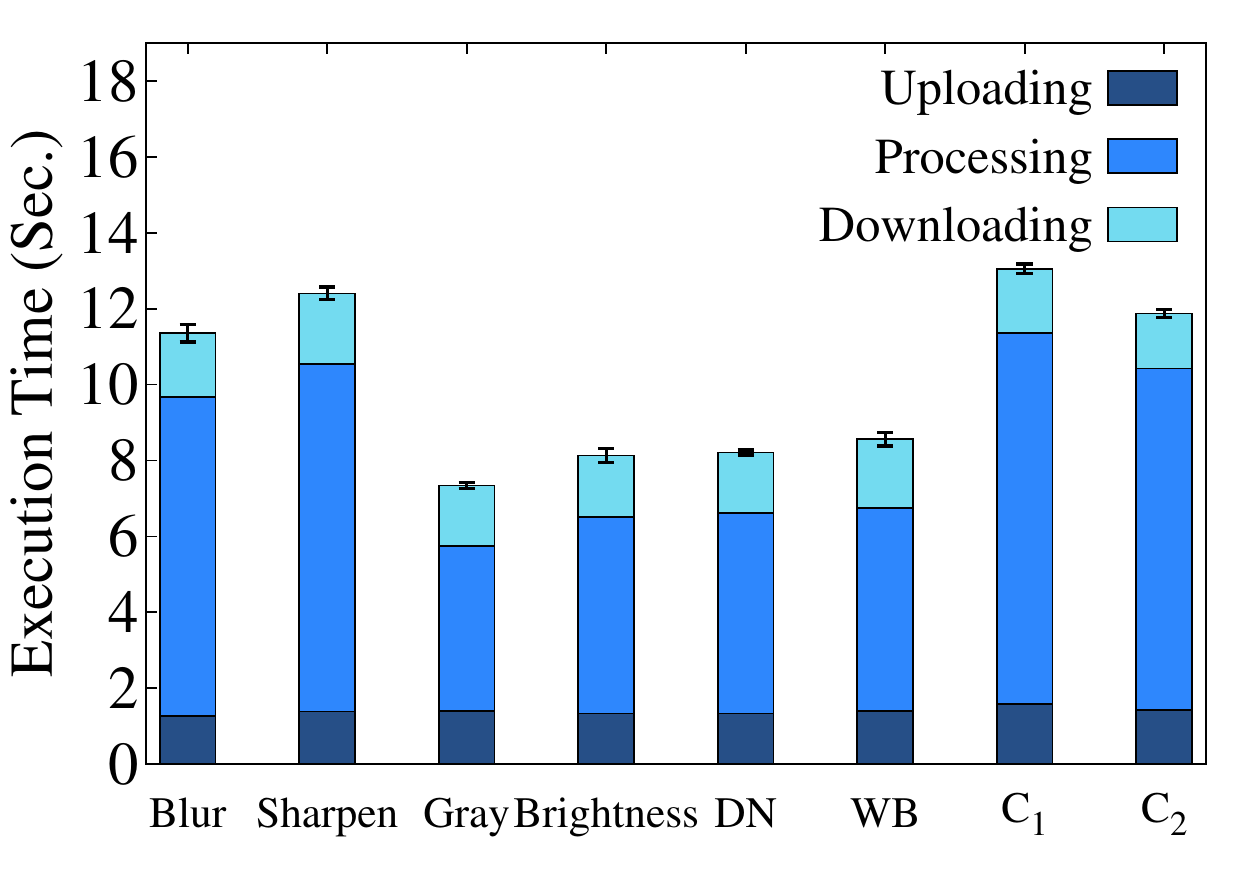}
    \caption{\em Execution time of \sname with different filters.}
    \label{fig:filters}
\end{figure}
\noindent\textbf{Overall execution time.}~~
Figure~\ref{fig:filters} shows the overall execution time of \sname for applying various filters to a 2-second 720p video.
The overall execution time includes the time to upload the video from the camera device, post-process it, 
and download it to the consumer. We show the results for applying a single filter (the first 6 bars) and for two-filter 
combinations (C1 is the sharpen filter followed by the white balance and de-noise filters, while C2 is the blur filter followed 
by the brightness and gray filters). Results show that \sname achieves quite decent performance: it takes about $7.3$ 
seconds to process the video with the most lightweight filter and about $13.1$ seconds for a heavyweight filter combination 
(C1). In contrast, as we show in \S\ref{sec:related}, prior systems take several orders of magnitude longer to achieve the same.

\begin{figure}
    \begin{minipage}[b]{0.3\columnwidth}
    \centering
    \includegraphics[width=1.8\columnwidth]{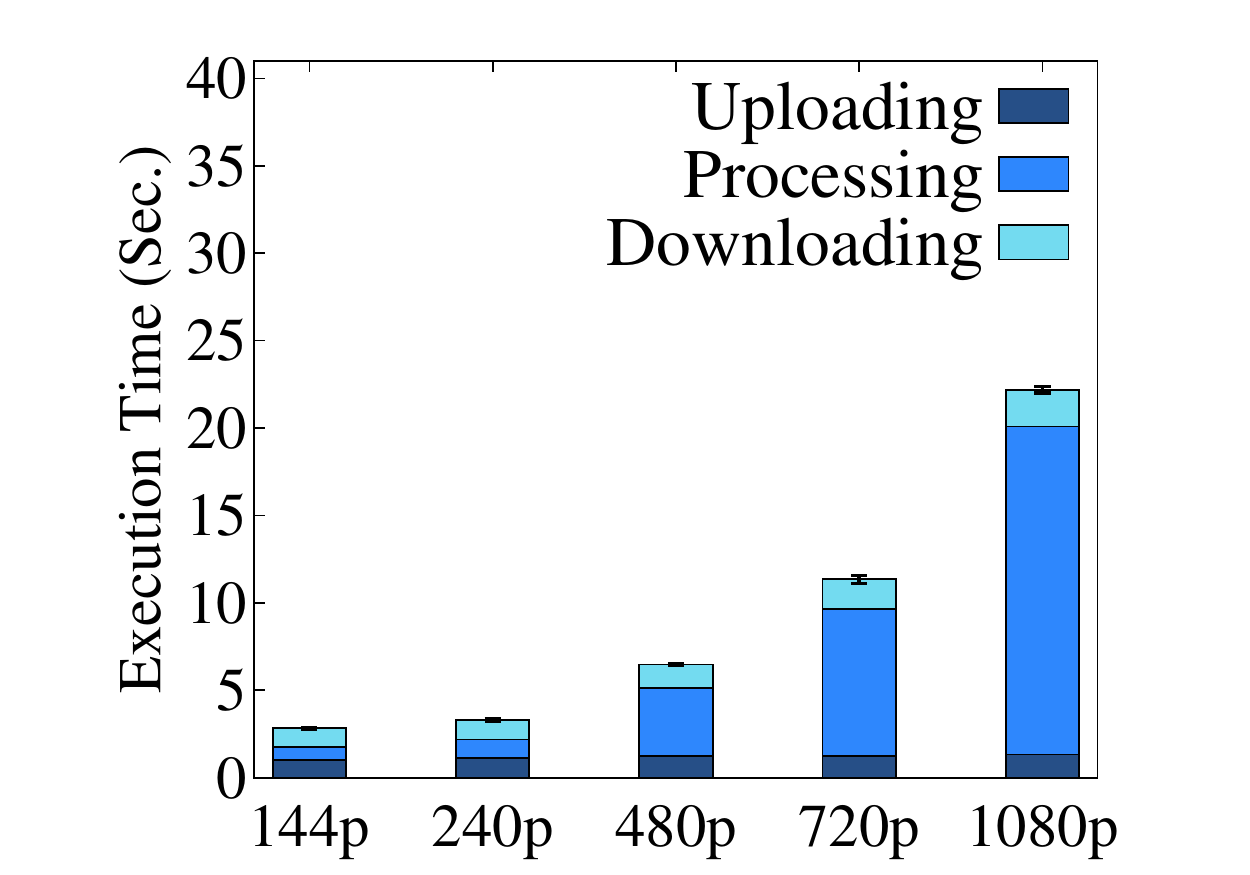}
    \end{minipage}
    \hspace{0.5in}
    \begin{minipage}[b]{0.3\columnwidth}
    \centering
    \includegraphics[width=1.8\columnwidth]{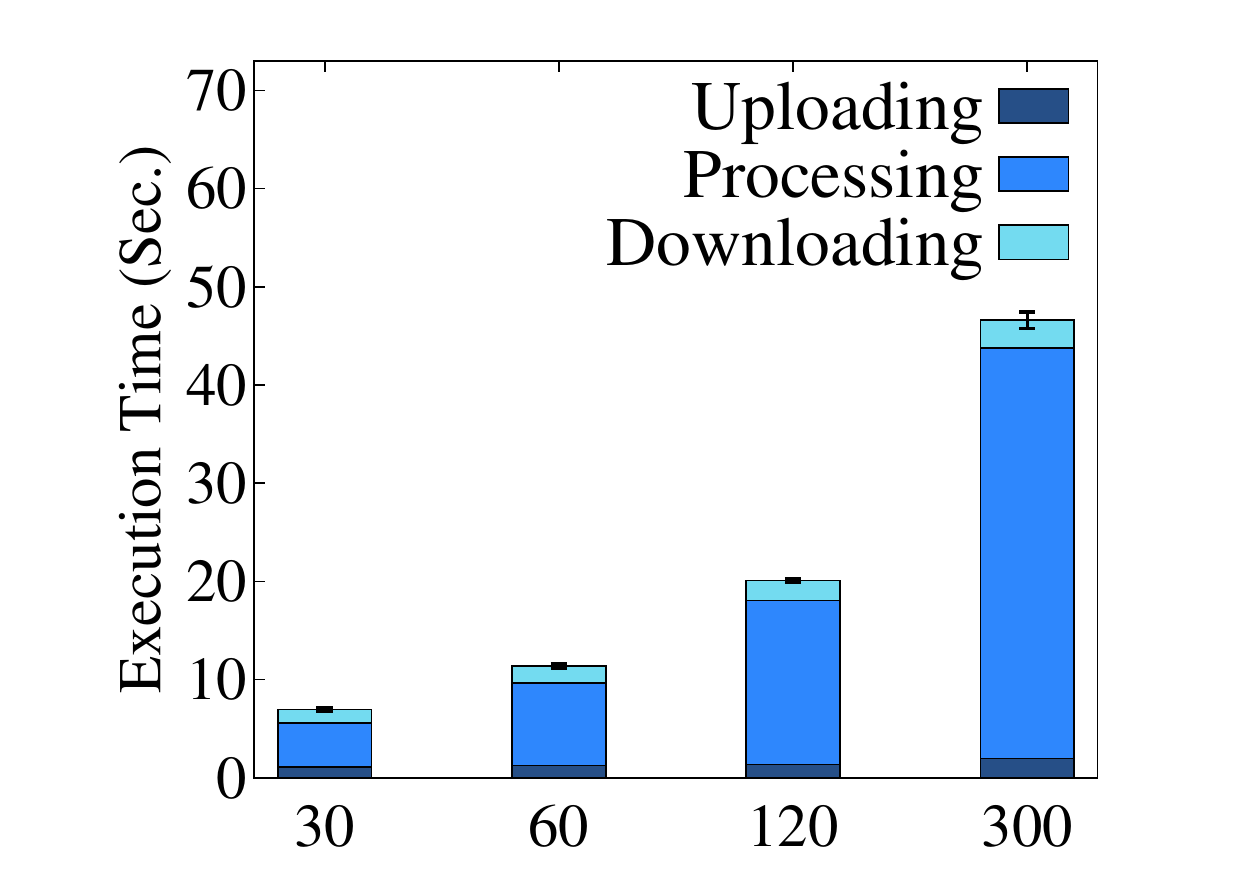}
    \end{minipage}
    \caption{\em Execution time of different resolution (Left) and different number of frames (Right).}
    \label{fig:performance_of_size_and_num}
    \vspace{-0.2in}
\end{figure}

We also measure overall execution time for different frame sizes or different numbers of frames.
Figure~\ref{fig:performance_of_size_and_num} shows the results, which show that execution time 
increases linearly with the growth in either frame sizes (in terms of the number of frame pixels) or the number of frames.

\noindent\textbf{The overhead of design decisions in \sname.}~~
As discussed in \S\ref{sec:design} and \S\ref{sec:multi}, two important \sname design choices are: hosting the 
post-processor in SGX enclaves and using fixed-function enclaves.
We now evaluate the overhead of each choice.
To do this, we introduce three baseline implementations of the video post-processor.
Baseline1 does away with the fixed-function enclaves and executes all the post-processor components in one SGX enclave.
Baseline2 does away with the use of enclaves (and digital signatures) and uses trusting OS processes 
to host different components of the post-processor. Baseline3 does away with both design decisions and 
executes all post-processor components in one OS process.
We also show the results for two-filter configurations: one with 2 filters (blur and sharpen) and one with 6 filters (all filters in 
the \sname prototype). For a fair comparison, we use the same number of threads for filters in all prototypes and use 
separate threads for decoding and encoding. We use 2 and 6 threads for filters for 2-filter and 6-filter experiments, respectively.

\begin{figure}
    \centering
    \includegraphics[width=0.8\columnwidth]{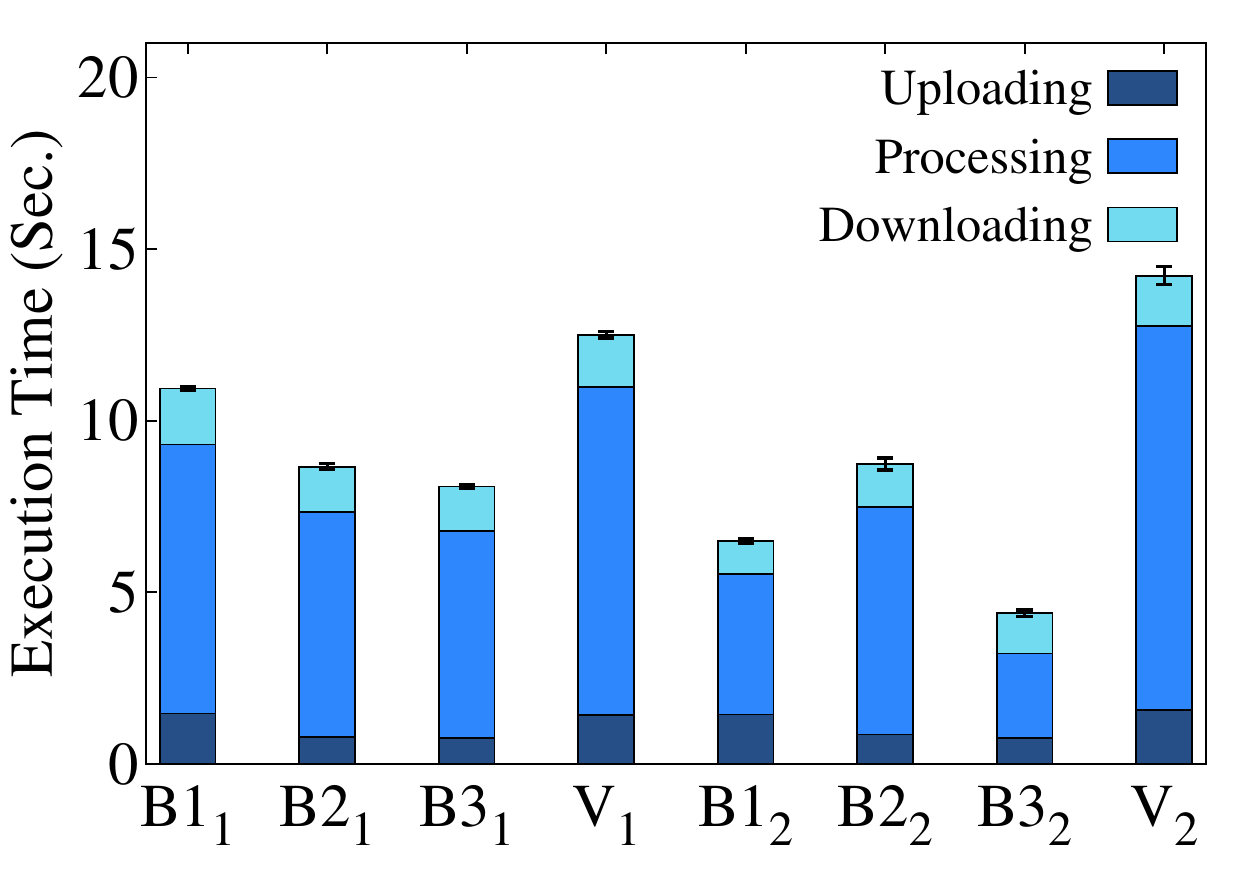}
    \caption{\em Execution time of different designs. B1, B2, B3, and V refer to Basline1, Baseline2, Baseline3, and \sname. 
    The last number in x-axis labels shows the configuration. 1 is the 2-filter configuration and 2 is the 6-filter configuration.}
    \label{fig:pipeline_bundle}
\end{figure}

The measurements are shown in Figure~\ref{fig:pipeline_bundle}.

The results of \sname vs. Baseline1 show that the use of fixed-function enclaves adds 14.3\% and 118.9\% performance 
overhead in the 2-filter and 6-filter configurations, respectively.
The major reason for much higher performance overhead in the second configuration is that each filter offers different performance. 
Note that ach filter runs as an individual thread, while, in Baseline1, all filters run together in one thread, 
though there are multiple  such threads.
This means that Baseline1 can make sure that all threads are making use of as much computing resource as possible, 
whereas, \sname may have some threads idling at times.
The results of \sname vs. Baseline2 show that the use of SGX enclaves adds 44.3\% and 62.8\% 
performance overhead in the 2-filter and 6-filter configurations, respectively.
Finally, the experiments (i.e., \sname vs. Baseline3) show that both design decisions together a
dd 54.7\% and 223.4\% performance overhead in the 2-filter and 6-filter configurations, respectively.
Given the benefits of these design decisions (i.e., the ability to provide verifiable provenance and ease of verification 
by the consumer), we believe that this overhead is acceptable.

Finally, note that Baseline1 and Baseline3 require some time to generate the post-processor binary, since the latter 
depends on the choice of filters. We measured compilation time from source to binary (assuming that the time to 
generate the source from existing filters is negligible): 6.5 and 4.1 seconds for Baseline1 and Baseline3, respectively. 

\begin{figure}
    \centering
    \includegraphics[width=0.5\columnwidth]{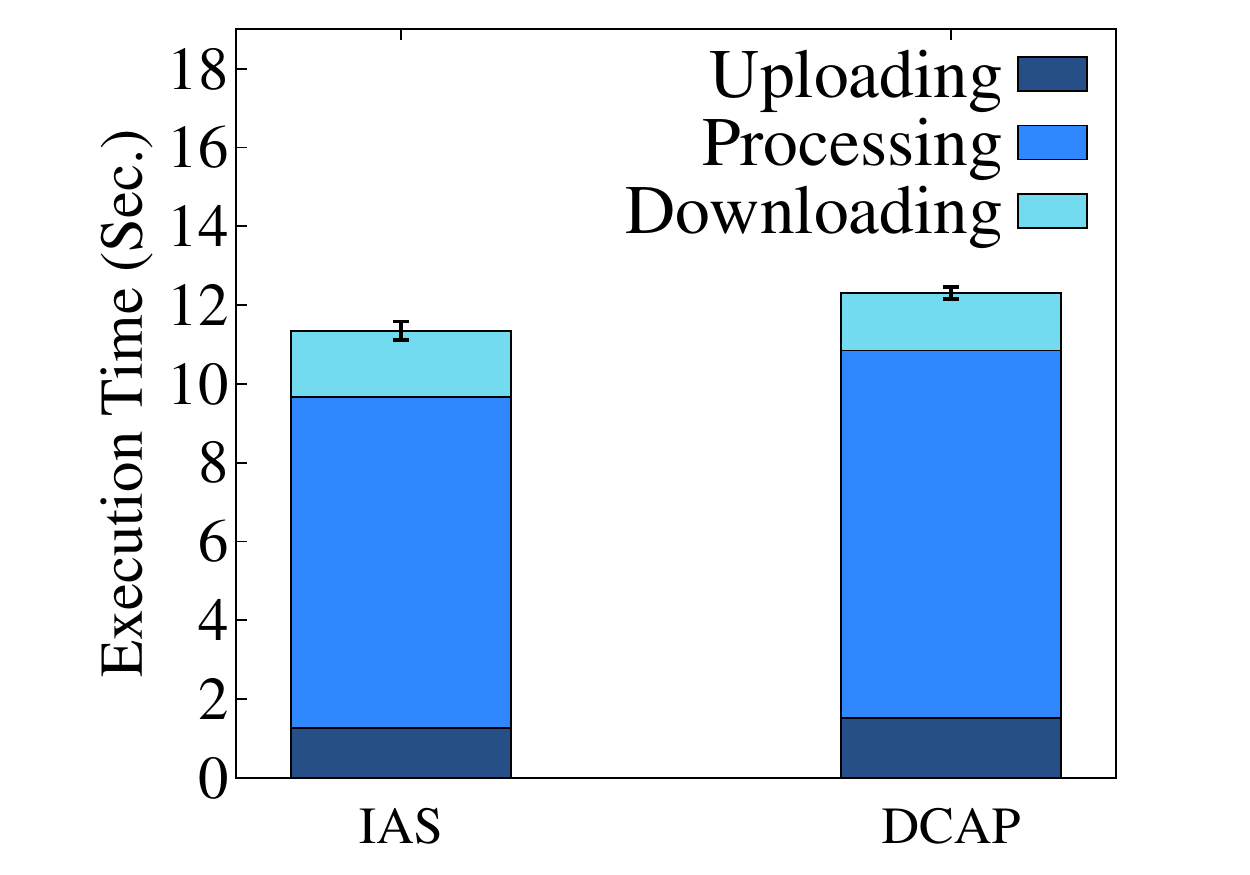}
    \caption{\em Comparison between IAS and DCAP based remote attestation.
	}
    \label{fig:ias_dcap_and_distributed}
    \vspace{-0.2in}
\end{figure}

\noindent\textbf{Comparison with YouTube.}~~
To get a sense of the usability implications of \sname's performance, we compare it with YouTube.
More specifically, we upload a 10-second video to YouTube and measure the time it takes for the video to be uploaded and processed by YouTube.
We suspect the processing includes transcoding the video.
It also includes several checks, such as copyright checks.
We use the same Galaxy S20+ smartphone under the same 5GHz Wi-Fi network for uploading the video
This takes about 31 seconds (with a maximum of 45 seconds and a minimum of 20 seconds), which compares favorably with \sname.
As Figure~\ref{fig:performance_of_size_and_num} (Right) shows, uploading and processing of the same 10-second video to \sname takes about 44 seconds.
Therefore, we believe \sname's performance is good and usable for offline use cases.

\noindent\textbf{Consumer.}~~
We measured the verification time for the consumer.
Results show that the consumer takes on average 3.4 seconds (with standard deviation of 0.0012 seconds) 
to verify the provenance information when SGX IAS and Android SafetyNet are used.
This shows that verification is fast enough for the consumers not to experience much playback latency.

\noindent\textbf{IAS vs. DCAP.}~~
We compare performance of \sname when using IAS or DCAP for remote attestation.
We use the same server that runs our enclaves to run the Provisioning Certificate Caching Service (PCCS) 
and Quoting Enclave (QE). Figure~\ref{fig:ias_dcap_and_distributed} shows the result.
We can see that, despite using a different remote attestation mechanism, video processing latency is 
roughly the same, with DCAP latency being slightly higher than that of IAS.
We hypothesize that this occurs because DCAP requires additional PCCS and QE software running along with the enclaves.

\noindent\textbf{Post-processing on a smartphone.}~~
As mentioned in \S\ref{sec:design}, we choose to perform video post-processing in powerful servers 
and workstations, rather than the camera device. Here, we justify this design choice empirically.
We use the same Galaxy S20+ camera device for this experiment.
By leveraging \emph{Android Native Development Kit} (NDK)~\cite{android_ndk}, we port the exact same filter codes to run in an Android app and measure their execution time.

The measurements show that the server spends 10.5 seconds to process frames using the blur filter.
This time includes the digital signature and provenance-related workload,  as well as uploading, downloading, 
decoding, and encoding. In contrast, Samsung Galaxy S20+ smartphone spends 32.4 seconds  just to apply the filter, 
which is over 3$\times$ slower than the server. Furthermore, note that S20+ is a powerful high-end device.
Weaker devices would take much longer to apply the same filters, e.g., Samsung Galaxy S8+ with Snapdragon 
835 takes 95.4 seconds, which is 9$\times$ slower than the server.

\section{Discussions \& Future Work}

\subsection{Performance Optimizations}
Our prototype achieves decent performance for off-line use-cases, where the captured video is not needed by the 
consumer immediately. However, it does not support real-time use-cases, notably, video conferencing.
We believe that \sname's design could be made suitable for real-time through further optimizations.
One way is to enable multi-threading in both decoder and encoder enclaves, and use multiple instances of the 
same filter enclaves. This way, multiple frames can be processed at the same time. 
Another possibility is to optimize performance of the filters themselves. As shown in the evaluation section, 
even baseline implementations cannot match real-time performance. 
To improve filter performance, we could enhance the CPU-based implementation or port it to run on accelerators, 
e.g., GPUs. Note that the latter would require TEE-enabled accelerators, e.g., Graviton~\cite{Volos2018}, which are not currently available.

\subsection{Camera Device Privacy}
\sname relies on the camera device to generate signatures over the captured video and its provenance info.
We assume that the camera TEE client holds a unique key-pair, the public part of which is signed by the camera vendor.
Usage of this public key might raise concerns regarding camera device privacy, since videos signed with the same 
private key can be linked. 
\sname's prototype addresses this by generating a fresh key-pair (each time a video is recorded) and provisioning it through 
SafetyNet. Albeit, the camera vendor must be trusted not to link the public keys to the camera device. 
Nevertheless, the ability to identify camera devices might be beneficial in certain use-cases, e.g., a consumer might 
prefer videos produced by reputable sources. However, the current \sname prototype does not support this.

\section {Related Work} \label{sec:related}

\begin{table*}[t] 
\centering
{\small
\begin{tabular}{|c|l|l|l|l|}
\hline
\multicolumn{1}{|l|}{} & \multicolumn{1}{c|}{\textbf{\begin{tabular}[c]{@{}c@{}}Granularity of Provenance\end{tabular}}} & \multicolumn{1}{c|}{\textbf{Performance (sec)}}                                          & \multicolumn{1}{c|}{\textbf{\begin{tabular}[c]{@{}c@{}}Videos?\end{tabular}}} & \multicolumn{1}{c|}{\textbf{Method}}                           \\ \hline
Alethia~\cite{aletheia}                & Coarse-grained                                                                                    & No measurements                                                                                      & Yes                                                                                       & \begin{tabular}[c]{@{}l@{}}Digital Signature\end{tabular}    \\ \hline
PhotoProof~\cite{Naveh2016}             & Fine-grained                                                                                      & 673.5 (128x128)                                                                          & No                                                                                        & \begin{tabular}[c]{@{}l@{}}Zero Knowledge Proof\end{tabular}  \\ \hline
YouProve~\cite{Gilbert2011}               & Coarse-grained                                                                                    & 28.0 (JPEG, 1296x972)                                                                    & No                                                                                        & \begin{tabular}[c]{@{}l@{}}Fidelity Analysis\end{tabular}    \\ \hline
FrameProv~\cite{Mansoor2019}              & Fine-grained                                                                                      & No prototype                                                                                      & Yes                                                                                       & \begin{tabular}[c]{@{}l@{}}Digital Signature\end{tabular}    \\ \hline
AMP~\cite{england2020amp}                    & Coarse-grained                                                                                    & 0.08 $\sim$0.24                                                                          & Yes                                                                                       & \begin{tabular}[c]{@{}l@{}}Fragile Watermarking\end{tabular} \\ \hline
\sname              & Fine-grained                                                                                      & \begin{tabular}[c]{@{}l@{}}4 (30 frames, 1280x720) \\ $\sim$40 (300 frames)\end{tabular} & Yes                                                                                       & \begin{tabular}[c]{@{}l@{}}Digital Signature\end{tabular}    \\ \hline
\end{tabular}
\caption{\em Comparison of Related Work.}
\label{tab:related}
}
\end{table*}

\noindent\textbf{Secure Video Provenance.}~~
Several video provenance methods have been proposed in the past~\cite{Gehani2007,Mansoor2019,Sorell2012,aletheia,england2020amp,Hasan2019}, including watermarking, hash chaining, use of digital signatures, and use of blockchain.
Table~\ref{tab:related} summarizes key related work and how they compare against \sname.
Granularity of provenance is defined as ``fine'' when the system provides information of how the video/photo was edited, and ``coarse'' otherwise.
Below, we describe the key related works in detail.

Some research focuses on providing provenance to a single photo~\cite{Naveh2016,Gilbert2011,Mirzamohammadi2020}.
PhotoProof~\cite{Naveh2016} uses zero knowledge proofs to cryptographically prove that the photo has been edited correctly without revealing the original photo.
However, the latency of generating and proving the proof is orders of magnitude slower compared to \sname.
Another related work that focuses on photo provenance is YouProve~\cite{Gilbert2011}.
Rather than giving fine-grained information on what kind of editing has been done, as \sname does, YouProve analyzes photos to understand where it was edited.

Some works extend photo provenance to provide provenance for a video.
One example is Aletheia~\cite{aletheia}, a tool that allows users to protect the provenance of their videos by signing them.
The main drawback of this system is that it requires consumers to trust the video without the ability to verify the filters applied to it.
Another example is FrameProv~\cite{Mansoor2019}, which provides provenance to a video of raw frames using a hashchain of frames.
In \sname, we do not blindly trust the system that conducts video processing and encoding, as FrameProv does.
Several related works have proposed a video provenance system that uses blockchain to prove and assess the provenance of the video~\cite{Hasan2019,amber,truepic}.
All of these systems suffer from high latency
because of their dependency on blockchain.
AMP~\cite{england2020amp} utilizes the Confidential Computing Framework~\cite{ccf} instead of blockchain to build a fast and reliable public ledger that keeps track of published media.
Anyone can view this ledger to verify the provenance of the media that they received.
AMP uses watermarking methods to embed provenance information to the media and therefore does not allow editing of the videos, as it breaks the watermarked information.
Moreover, the trust of the media depends on the trust of the publisher.

Media provenance is gaining attraction not only in academia but also in the industry.
An early attempt on trustworthy photo capture in industrial setting is Canon's Original Data Security Kit~\cite{canon_osk} to prove image originality.
However, later on it was proven that there was a flaw in the system~\cite{elcomsoft_osk}.
Shifting our eyes to present time, startup companies such as Truepic~\cite{truepic} and Serelay~\cite{serelay} aim to provide verifiable provenance to photos using TEEs.
The two products currently only provide trusted photo capturing.
Moreover, we see their trusted cameras as complementary to our system, as it will provide users with a wide variety of trusted camera devices to choose from.
In addition, iPhones uses a physically isolated Secure Enclave Processor to securely control the camera~\cite{apple_sep_certification}.
While not the case today, the same design can be used to provide secure provenance as well.

Both Truepic and Serelay follow the Content Authenticity Initiative (CAI)~\cite{cai} standard for their provenance info.
CAI is an organization that aims to establish a standard in media provenance.
CAI's vision is aligns with \sname; providing \emph{fine-grained} provenance information at an \emph{``reasonable'' performance}.
Since CAI is a standard, it cannot be directly compared to \sname.
However, we believe \sname is a step towards achieving CAI's requirements for videos.

\noindent\textbf{System Provenance Technologies.}~~
Linux Provenance Module (LPM)~\cite{Bates2015} and CamFlow~\cite{Pasquier2017} are technologies that allow capture of system-wide provenance information.
Although these works and \sname share the interest of capturing verifiable provenance information of certain data, the trust model is different.
Namely, while LPM and CamFlow requires OS and kernels to be trusted, \sname assumes that the entire system, including priviledged software, to be untrusted.

\noindent\textbf{Secure Video Processing using GPU TEEs.}~~
There has been several research done to realize TEEs that run on top of GPUs, notably Graviton~\cite{Volos2018} and Telekine~\cite{Hunt2020}.
Visor~\cite{Poddar2020} is a secure, privacy-preserving video analytics platform that is built on top of Graviton (which is an proprietary emulator).
It provides confidentiality to the video analytics pipeline and protects the video data from attacks including side-channel attacks.
In contrast, in \sname, we are not concerned with confidentiality of the video.
We envision utilizing GPU TEEs in \sname in the future to provide more complex filters such as edge detection and object recognition.

\section{Conclusions}
\label{sec:conclusions}
We envision a world where consumers can easily verify authenticity and integrity of videos 
using trustworthy provenance information. Current video veracity techniques are either too 
coarse-grained in provenance (i.e., prove little to the consumer), or offer poor performance -- a matter
of minutes for a small single frame.

\sname provide precise provenance information on both the original video recording as well as  
every step of post-processing. \sname defends against attackers that control all network communication 
as well as all software state outside TEEs. The prototype described in this paper can be immediately 
deployed and used. Its evaluation on Intel SGX enclaves in Azure Confidential Computing VMs shows 
that \sname's performance is good and usable for offline use cases and is 
orders of magnitude 
faster than the current techniques.

\section*{Acknowledgments}

The work was supported in part by NSA NCAE-C Cyber Curriculum and Research 2020 Program Award \#H98230-20-1-0345 and UCI ICS Exploration Research Award.

\balance
\footnotesize
\raggedright
\bibliographystyle{abbrv}
\bibliography{paper}

\end{document}